\documentclass[runningheads]{llncs}
\usepackage{tabularx, makecell, booktabs}
\usepackage{alltt}
\usepackage{caption}
\usepackage{paralist}
\usepackage{nicefrac}
\usepackage{booktabs}
\usepackage{algorithm}
\usepackage{algorithmic}
\usepackage{xspace}
\usepackage{amsmath}
\usepackage{svg}
\usepackage{marvosym}
\usepackage{wasysym}
\usepackage{verbatim}
\usepackage{hyperref}
\usepackage{multirow}
\usepackage{cite}
\usepackage[capitalize]{cleveref}
\usepackage[per-mode=symbol,detect-all]{siunitx}
\usepackage{graphics}
\usepackage{amssymb}
\usepackage{wrapfig}
\usepackage{enumitem}
\usepackage{placeins}
\usepackage{tabularx}
\usepackage{rotating}
\usepackage{listings}
\usepackage{xcolor}
\usepackage{adjustbox} 

\usepackage{tikz,ifthen,pgfplots}
\usepackage{graphicx}
\usepackage{svg} 
\usepackage{graphicx}
\usepackage{import} 
\usepackage{arydshln}
\usepackage{tcolorbox}
\tcbuselibrary{listings,skins}

\usepackage{listings}
\usepackage{xcolor}

\usepackage{caption}
\usepackage{subcaption}

\usepackage{adjustbox}
\usepackage{setspace}
\lstdefinestyle{codestyle}{
	basicstyle=\ttfamily\small,
	frame=single,
	breaklines=true,
	columns=fullflexible,
	keepspaces=true
}

\lstdefinelanguage{smtlib}{
	morekeywords={set-logic,declare-fun,assert,check-sat,exit,define-fun},
	sensitive=true,
	morecomment=[l]{;},
}

\lstdefinestyle{asmstyle}{
	label={lst:SMT-LIB-code},
	basicstyle=\ttfamily\scriptsize,
	frame=single,
	breaklines=false,          
	keepspaces=true,           
	columns=fullflexible,      
	xleftmargin=0pt,
	xrightmargin=0pt,
	framexleftmargin=0pt,
	framexrightmargin=0pt,
	aboveskip=2pt, 
	belowskip=2pt,
	tabsize=1,                 
	numbers=none,
	showstringspaces=false
}

\lstdefinestyle{libstyle}{
	label={lst:SMT-LIB-code},
	basicstyle=\ttfamily\scriptsize,
	frame=single,
	breaklines=true,          
	keepspaces=true,           
	columns=fullflexible,      
	xleftmargin=0pt,
	xrightmargin=0pt,
	framexleftmargin=0pt,
	framexrightmargin=0pt,
	aboveskip=2pt, 
	belowskip=2pt,
	tabsize=1,                 
	showstringspaces=false
}

\usetikzlibrary{arrows,trees,backgrounds,automata,shapes,decorations,plotmarks,fit,calc,positioning,shadows,chains}

\lstdefinestyle{mystyle}{
	basicstyle=\ttfamily\footnotesize,
	backgroundcolor=\color{lightgray!20},
	frame=single,
	keywordstyle=\color{blue},
	commentstyle=\color{green!60!black},
	stringstyle=\color{red!60!black},
	numberstyle=\tiny\color{gray},
	breaklines=true,
	numbers=left,
	stepnumber=1,
	numbersep=5pt,
	tabsize=4,
	showspaces=false,
	showstringspaces=false,
	showtabs=false
}

\definecolor{color0}{RGB}{228,87,46}
\definecolor{color1}{RGB}{23,190,187}
\definecolor{color2}{RGB}{255,201,20}
\definecolor{color3}{RGB}{46,40,42}
\definecolor{color4}{RGB}{118,176,65}

\definecolor{color0}{RGB}{250,121,33}
\definecolor{color1}{RGB}{254,153,32}
\definecolor{color2}{RGB}{185,164,76}
\definecolor{color3}{RGB}{86,110,61}
\definecolor{color4}{RGB}{12,71,103}

\definecolor{color0}{RGB}{228,253,225}
\definecolor{color1}{RGB}{138,203,136}
\definecolor{color2}{RGB}{100,131,129}
\definecolor{color3}{RGB}{87,87,97}
\definecolor{color4}{RGB}{255,191,70}

\definecolor{color0}{RGB}{226,59,62}
\definecolor{color1}{RGB}{243,114,44}
\definecolor{color2}{RGB}{248,150,30}
\definecolor{color3}{RGB}{249,199,79}
\definecolor{color4}{RGB}{126,179,86}
\definecolor{color5}{RGB}{67,170,139}
\definecolor{color6}{RGB}{39,125,161}
\definecolor{color7}{RGB}{21,49,60}
\definecolor{color8}{RGB}{180,215,228}

\tikzstyle{every pin edge}=[<-,shorten <=1pt]
\tikzstyle{neuron}=[circle,fill=black!25,minimum size=17pt,inner sep=0pt]
\tikzstyle{input neuron}=[neuron, fill=green!50]
\tikzstyle{output neuron}=[neuron, fill=red!50]
\tikzstyle{hidden neuron}=[neuron, fill=blue!50]
\tikzstyle{annot} = [text width=4em, text centered]
\usetikzlibrary{calc}

\newcommand{\mysubsection}[1]{\medskip\noindent\textbf{#1}}

\definecolor{nnedgecolor}{RGB}{90,90,90}
\tikzstyle{every pin edge}=[<-,shorten <=1pt]
\tikzstyle{every path}=[draw=color7!50]
\tikzstyle{neuron}=[circle,fill=black!25,minimum size=17pt,inner sep=0pt]
\tikzstyle{input neuron}=[neuron, fill=color4]
\tikzstyle{output neuron}=[neuron, fill=color0]
\tikzstyle{hidden neuron}=[neuron, fill=color6!80]
\tikzstyle{annot} = [text width=4em, text centered]
\tikzstyle{nnedge} = [-{stealth},shorten >=0.1cm, shorten <=0.05cm,line 
width=0.8pt,nnedgecolor]
\usetikzlibrary{calc}

\begin{document}

\title{ReSMT: An SMT-Based Tool\\for Reverse Engineering}

\author{Nir Somech and Guy Katz }
  
\institute{The Hebrew University of Jerusalem, Jerusalem, Israel\\
  \email{nirsomech@cs.huji.ac.il, g.katz@mail.huji.ac.il}
}

\maketitle

\begin{abstract}
	
  Software obfuscation techniques make code more difficult
  to understand, without changing its functionality.  Such techniques
  are often used by authors of malicious software to avoid
  detection~\cite{Lu2024}.  \emph{Reverse Engineering}
  of obfuscated code, i.e., the process of overcoming obfuscation and
  answering questions about the functionality of the code, is
  notoriously difficult; and while various tools and methods exist for
  this purpose, the process remains complex and slow, especially when
  dealing with layered or customized obfuscation techniques.
         
  Here, we present a novel, automated tool for addressing some of the
  challenges in reverse engineering of obfuscated code. Our tool,
  called ReSMT, converts the obfuscated assembly code into a complex 
  system of logical assertions that represent the code functionality, 
  and then applies SMT solving and simulation tools to inspect the 
  obfuscated code's execution. The approach is mostly automatic, 
  alleviating the need for highly specialized deobfuscation skills. 
  In an elaborate case study that we conducted, ReSMT successfully 
  tackled complex obfuscated code, and was able to solve reverse-engineering 
  queries about it. We believe that these results showcase the potential 
  and usefulness of our proposed approach.
	
\end{abstract}

\section{Introduction}
\label{sec:introduction}

\begin{figure}[!htbp]
	\centering
	\begin{subfigure}[t]{0.45\textwidth}
          \centering
		\includegraphics[height=4cm, keepaspectratio]{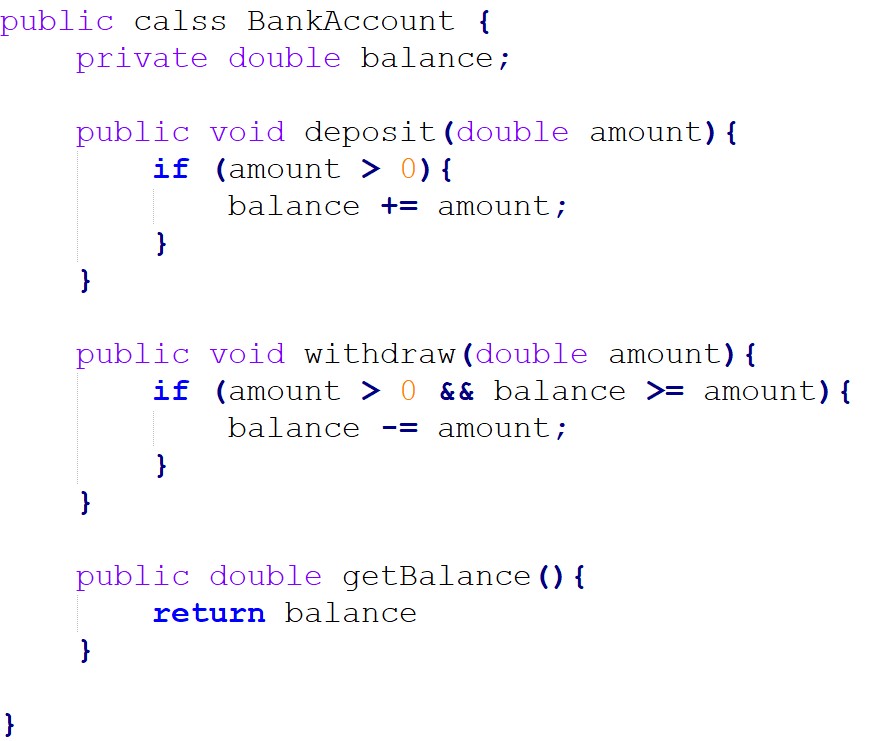}
		\caption{Original code.}
	\end{subfigure}
	\hfill
	\begin{subfigure}[t]{0.45\textwidth}
          \centering
		\includegraphics[height=4cm, keepaspectratio]{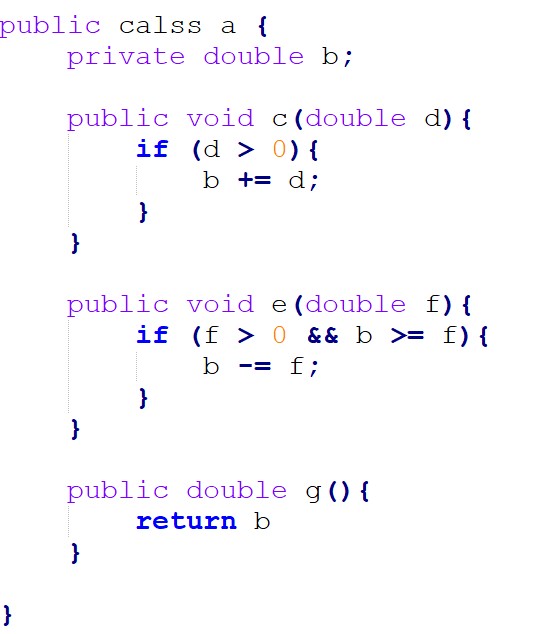}
		\caption{Obfuscated code.}
	\end{subfigure}
	
	\caption{Example of java function code obfuscation: original code
		(left) vs. obfuscated code (right).}
	\label{fig:obfuscation_example}
\end{figure}

In software development, obfuscation is the practice of making source
code or machine code difficult for humans to read and understand
\cite{BeLa15}. Obfuscation techniques serve as a primary defense
mechanism against hacking attempts and help protect against common
attacks such as code injection, reverse engineering, and tampering
with users' personal information. Fig.~\ref{fig:obfuscation_example}
depicts a simple example of Java code and an obfuscation thereof.

Unfortunately, code obfuscation is often applied in malware, to make
it harder for security researchers to analyze and understand its
logic. This delays the reverse-engineering process --- the process of
examining compiled code and figuring out its various properties ---
making it more difficult to develop security logic and protective
mechanisms to counteract the malware \cite{Lu2024}. One possible path
is to restore the obfuscated code to its original, comprehensible form
so that it can be inspected; but this process, known as
\emph{deobfuscation}, is often difficult, time-consuming, and requires
significant manual effort. Although various automated deobfuscation
approaches have been proposed, they are often brittle, have limited
scope, and fail to produce comprehensive results~\cite{Jj2025, Bh2024}. 
Given these limitations and the widespread use of diverse obfuscation 
techniques by malware and malicious actors, it is essential to develop 
effective tools for automated reverse engineering.
See Appendix~\ref{Deobfuscation_appendix} for additional information 
on deobfuscation methods.

In this paper we present ReSMT: an automated, SMT-based revsere
engineering framework. At a high-level, ReSMT operates by converting
the assembly code being inspected to an intermediate language, such
that every assembly instruction is represented using a small set of
atomic instructions. These instructions are then converted into a set
of constraints and assertions, which constitutes a logical formula.  To
this formula we then add constraints that represent the
reverse-engineering goal that we aim to resolve.  Finally, an SMT
solver backend is used in order to search for a satisfiable solution
for the query.  If the query is determined to be satisfiable, the
assignment found by the SMT solver describes an execution of the
program (i.e., indicates the value of the different registers, memory
values, and arguments) that satisfies the reverse-engineering goal.

Delving deeper, as the first step of its operation, ReSMT analyzes
compiled byte code and disassemble it into the more comprehensible
form of assembly instructions.  In the second step, ReSMT converts the
assembly instructions into instructions in a high-level intermediate
language (IL), from which a logical formula will later be extracted.
Here, a major task was to accurately capture the exact semantics of the
assembly instructions and their inter-relations, in order to ensure
soundness. To address this key challenge, we harnessed the
OpenREIL framework~\cite{DuPo}, designed to convert assembly
instructions into a platform-independent intermediate language.  This
was a significant engineering challenge, as the OpenREIL framework has
not been maintained for 9 years, and its functionality and key
components lacked documentation. Further, it uses many sub-libraries
that are themselves no longer maintained. Eventually, however, we were
able to create an up-to-date, operational version of the OpenREIL
core, and integrate it into ReSMT.

The third step of our pipeline is to receive OpenREIL's output
(namely, instructions in the platform-independent IL) and transform it
into a set of logical assertions in SMT-LIB format. As part of this
process, one must consider how each IL instruction interacts with the
underlying computer architecture: memory, registers, stack, heap, and
other crucial subsystems; and each of these interactions must also be
expressed as logical assertions in SMT-LIB. To perform this, we
integrated into ReSMT additional libraries that implement a
symbolic-execution engine on top of OpenREIL's emulator, and which
allowed us to properly capture the aforementioned interactions.
Another important task that occurs in this step is the addition of the
RE goal we aim to solve as assertions in the logical formula.

The fourth and final step of the ReSMT pipeline is to dispatch the
SMT-LIB formula generated in the previous step using an off-the-shelf
SMT solver. In our evaluation we used Z3~\cite{DeBj08}. A SAT answer
by the solver typically indicates an input and an execution of the
program that triggers the behavior sought after as part of the RE
goal.

To evaluate ReSMT, we conducted a case study with six different
programs, which are similar in functionality but which vary widely
in the obfuscation techniques applied to them. The first version was
the original, unobfuscated program; and from it we generated five
obfuscated variants, by incrementally applying additional layers of
obfuscation. The reverse-engineering goal for all programs was to
discover the specific input which causes the program to successfully
complete its operation (the ``secret key'').  Thus, if the underlying
solver determined that the query was satisfiable, it also revealed the
specific input value that would produce the desired output.  While all
program variants were identical in logic and functionality, the final
variants, which were obfuscated most heavily, posed a significant
challenge to a human engineer trying to deobfuscate them manually.
Still, ReSMT was able to tackle all obfuscated programs, demonstrating
strong performance in terms of runtime and efficiency. We believe that
these results showcase the significant potential of our approach.

To summarize, our work has two main contributions:
\begin{inparaenum}[(i)]
  \item
 a new
tool, which can be used to assist in analyzing various obfuscated
binaries, including malware; and
\item
 an extensive
case-study, in which our new tool demonstrates very favorable results,
highlighting its potential.
\end{inparaenum}

The rest of the paper is organized as follows.
Section~\ref{sec:background_obfuscation} contains background on code
obfuscation, reverse engineering and SMT solving. In
Section~\ref{sec:our_approach} we present ReSMT's underlying approach
for automatically reverse-engineering obfuscated code.
Next, in Section~\ref{sec:featuresAndDesign} we discuss the features
and design of ReSMT, followed by a description of our case study in
Section~\ref{sec:case_study}.  Related work is covered in
Section~\ref{sec:Related_Work}, and we conclude and discuss future
work in Section~\ref{sec:Discussion}.

\section{Background}
\label{sec:background_obfuscation}

\mysubsection{Obfuscation.}  Software obfuscation refers to making code
difficult for humans to read and understand, while retaining its
original functionality or behavior
\cite{Lu2024}. Specifically, it changes how code is
presented and delivered, making reverse engineering and analysis
challenging. Obfuscation is commonly used to protect intellectual
property, hide program logic and sensitive values, and defend against
unauthorized code access. However, it also increases code size and can
degrade performance.
  
\begin{wrapfigure}[13]{r}{0.5\textwidth}
	\centering
	\includegraphics[width=\linewidth,keepaspectratio]{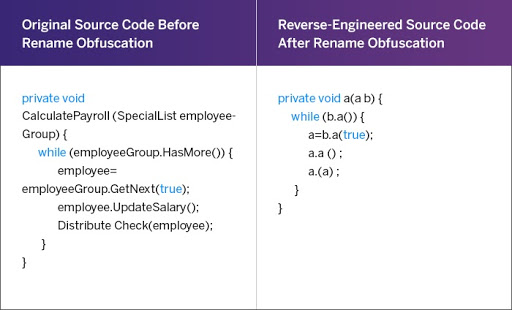}
	\caption{Original source code, and its obfuscated counterpart,
          after renaming.}
	\label{fig:rename_obfuscation_example}
\end{wrapfigure}

Obfuscation techniques are often combined to create layered
defenses. Common methods include \textit{renaming} variables and
functions to meaningless names (see
Fig.~\ref{fig:rename_obfuscation_example} for an example),
\textit{breaking control flow} into nonlinear sequences,
\textit{instruction pattern transformation} using complex equivalents,
\textit{dummy code insertion}, \textit{string encryption}, and
\textit{code transportation} by reordering
blocks~\cite{Lu2024, BeLa15, Ba2025}. Malwares often 
employ additional techniques such as \emph{XOR encoding}, which hides 
stringsand code via a bitwise exclusive OR operation with a key, and
\emph{ROT-13 substitution}, used to further obscure content~\cite{Lu2024, YoYi2010}.  
Numerous academic studies have surveyed existing obfuscation
techniques~\cite{Ga20, LyMiZhXu2020, ScKaKiMeWe2016, KuLa2015}; and there 
are multiple tools available that implement these techniques (e.g.,
ProGuard \cite{guardsquare}, Jscrambler \cite{jscrambler}, Dotfuscator
\cite{preemptive2024a}, VMProtect \cite{VMP} and Themida \cite{TM}).

\mysubsection{Reverse Engineering.}
\emph{Reverse engineering} (RE) is typically defined as the process 
of analyzing a product, device or system to understand its design,
construction or functionality. It involves disassembling, examining
and studying the components and structure of an existing object to
create a detailed representation or model of it~\cite{Ei05, Ne1996}. The
primary goal of reverse engineering is to extract useful information
or knowledge about how something works or is constructed without
access to its original design documentation or specifications.

Our scope in this work is the reverse engineering of obfuscated binary
code. Our straightfoward way to achieve this is to deobfuscate the
code in order to render it comprehensible, and then manually 
inspect it. However, deobfuscation is notoriously difficult: if done
manually, it requires a highly specific skill-set, and significant
time; and while there exists automated deobfuscation techniques, these
are often efficient only against the specific obfuscation techniques
they were designed to handle, but may fail when the developers of the
obfuscated code apply non-standard techniques. See
Appendix~\ref{Deobfuscation_appendix} for additional details.

From our perspective, deobfuscation is not the ultimate goal; rather,
it is merely a means for extracting some useful information about the
program. Broadly, given an obfuscated program $P$ that transforms
input $x$ to output $y$, we define a reverse-engineering condition
$\phi(y)$ to be a predicate on the outputs of the program, which
evaluates to true only when a certain condition is met; and our goal
is to find an input $x$ for which the formula $y=\phi(P(x))$ is
true. In other words, our goal is to find an input $x$ that correctly
triggers the desired output behavior. For example, if an obfuscated
program functions as a safe that only opens (i.e., prints some
output) when a specific input is entered, we might specify $\phi(y)$
to indicate that the desirable output has been printed, and then seek
the right $x$.
    
\mysubsection{SMT Solvers.}
Satisfiability Modulo Theories (SMT) is the problem of deciding
whether a given first-order formula is satisfiable, while taking into
account the rules of a background theory. This background theory
specifies certain restrictions on models that can satisfy the formula,
in the form of additional formulas that they must also satisfy. Common
background theories include arithmetic over integers and reals,
bit-vectors, arrays, or strings.  SMT can be viewed as a natural
extension of the Boolean Satisfiability Problem (SAT) to richer
logics: whereas SAT deals only with propositional logic, SMT formulas
can more readily encode the behavior of complex computer programs. The
SMT problem is in general quite difficult, although efficient decision
procedures exists for some background theories\cite{BaSeSeTi09}.

SMT-LIB \cite{CaFoTi16} is a language and standard for
formulating SMT queries. SMT-LIB allows for cross-solver compatibility
and is widely used in formal verification, symbolic execution, and
constraint solving tasks.

\section{Our Approach: An SMT-Based Solution}
\label{sec:our_approach}

In order to automate the reverse engineering process, ReSMT seeks to
transform the problem into an SMT query, typically involving the
theories of bit-vectors and arrays, which can then be dispatched using
an off-the-shelf solver. In order to construct this query, ReSMT
parses the obfuscated code and extracts from it instructions in an
intermediate language (IL). In turn, these instructions
are transformed into logical assertions that represent the code
execution, as well as the the different components of the computer
architecture during this execution (memory, stack, heap, registers,
etc.). To this query we also add assertions that represent the RE goal
we are trying to solve.
If the SMT solver succeeds in finding a satisfying assignment, it
constitutes an execution of the program that achieves the RE
goal. Conversely, if the SMT solver returns UNSAT, there is no way to
solve the RE question for the provided program, with the specified 
program and configuration. This could indicate that the RE goal that
was posed is unattainable (e.g., there are no inputs for which certain
outputs are produced), and that it may need to be refined. 
Finally, the SMT solver might also time-out without providing a conclusive
answer.

More concretely, our approach has four main steps: bytecode
disassembly, converting assembly code to an intermediate language,
converting the intermediate language into a logical formula, and
dispatching the formula using an SMT solver. Below we elaborate on
these steps and demonstrate them on a toy example of a code for
computing and returning the sum of two numbers (on the x86 CPU architecture).

\mysubsection{Step 1: Bytecode Disassembly.}
The first step in the process is extracting the function bytecode and 
disassembling it back to assembly instructions. This step is required
in order to ``make sense'' of the compiled code, so that it can later
be formulated as an SMT query.
This task, which is similar to the one performed by modern disassemblers,
is quite challenging. For instance, it requires taking into account
the CPU architectures, e.g., x86, x64, ARM, or ARM64 --- each of which
has a distinct instruction set, register configuration, and
hardware-specific constraints.
Additionally, various security mechanisms may be applied to the bytecode 
to hinder disassemblers, making it more difficult to reconstruct readable 
and accurate assembly instructions.

We illustrate the disassembly process through an example. Consider the 
bytecode on the left, and its corresponding assembly instructions on the 
right. The assembly code performs addition between two numbers and returns 
to the calling function. The first instruction assigns the value stored 
in the $ecx$ register to the $eax$ register; and the second instruction 
adds to $eax$ the value stored in $edx$ register. 
Finally, the $ret$ instruction pops the return address from the stack and 
assigns this value to the instruction pointer $eip$.  

\begin{figure}[htbp]
	\centering
	\begin{adjustbox}{width=0.7\linewidth}
		\begin{minipage}{\linewidth}
			\captionsetup{justification=centering}
			\label{fig:basic_bytecode}
			\vspace{0.3em}
			\begin{lstlisting}[language={[x86masm]Assembler},basicstyle=\ttfamily\footnotesize,frame=single]
				89 C8 01 D0 C3  ---->  mov eax, ecx
				                       add eax, edx
				                       ret
			\end{lstlisting}
		\end{minipage}
              \end{adjustbox}
              \caption{Simple bytecode and its corresponding assembly
                instructions.}
              \label{fig:exampleStep1}
\end{figure}

Existing, well-known tools for binary disassembly include IDA~\cite{HexRays2024}, objdump~\cite{objdump}, Windbg~\cite{windbg}, 
and others.

\mysubsection{Step 2: Converting the Assembly Code to an Intermediate
  Language.}  In the second step we convert each assembly instruction
into one or more instructions in an intermediate language (IL). The
motivation for this step is that assembly instructions are
complex, in the sense that a single action may have many effects; whereas IL
instructions are simpler, and so each can later be translated into
a single assertion or term for the SMT solver. Thus, each IL
instruction either describes the core action of the source assembly
instruction, or ``side-effects'' that directly arise from that
instruction (e.g., the setting of CPU flags or memory value
changes). At this point, we assume that sets of IL instructions
originating from different assembly instructions do not interact.

Another motivation for using IL is the multitude of different CPU
architectures, and the corresponding multitude of CPU instruction
sets. Adding the intermediate language layer, into which different
CPU instruction sets are translated, adds modularity to our design ---
making it fairly straightforward, even if time-consuming, to add
support for additional CPU architectures. In ReSMT, we used the
Reverse Engineering Intermediate Language (REIL) as our intermediate
language. REIL was designed specifically for the purpose of
transforming assembly code into higher-level code, more amenable to
analysis.

REIL~\cite{DuPo} has a reduced instruction, containing just 17
different instructions --- far fewer than those supports by common CPU
architectures, such as X86, ARM or ARM64. Still, it supports these
three architectures, and can translate many of their native
instructions into sets of REIL instructions. The REIL architecture is
register-based, without an explicit stack and with a flat memory
model. Further, it provides an unbounded number of registers.
The IL instruction sets can be divided into the following sets:
\begin{itemize}[topsep=0pt, partopsep=0pt, itemsep=0pt, parsep=0pt, leftmargin=1.5em]
	\item \textbf{Arithmetic Instructions}: addition, subtraction, 
	unsigned multiplication, unsigned division, unsigned modulo and 
	logical shift operations.
	\item \textbf{Bitwise Instructions}: bitwise AND, bitwise OR and bitwise 
	XOR operations.
	\item \textbf{Data Transfer Instructions}: load a value from memory, 
	store a value to memory and store a value in a register operations.
	\item \textbf{Conditional Instructions}: compare a value to zero and 
	conditional jump operations.
	\item \textbf{Other Instructions}: undefine a value, an ``unknown source''
	instruction and a ``no operation'' operations.
\end{itemize}
\vspace{-0.5em} 
\vspace{1em} 

Despite its expressiveness, the current version of REIL does not
support the full instruction set of any of the supported architecture.
For example, it cannot handle FPU instructions, or the MMX and SSE
extensions of x86; nor can it handle system calls, interrupts and
kernel-level instructions. Extending REIL to support these operations
is left for future work.

To continue our running example, Fig.~\ref{fig:exampleStep2} depicts 
the assembly instructions from Fig.~\ref{fig:exampleStep1}, and their 
corresponding encoding in IL.
The first assembly instruction \texttt{mov eax, ecx} is encoded to 
the IL instruction \texttt{STR	 R\_ECX:32,  R\_EAX:32}, meaning 
that the value of the 32-bit register\texttt{ecx} is to be stored 
in the 32-bit register \texttt{eax}. The second assembly instruction 
\texttt{add eax, edx} is encoded as the IL instruction 
\texttt{ADD	 R\_EAX:32, R\_EDX:32, R\_EAX:32}, meaning that the value 
presently in the \texttt{eax} register should be increased by the 
value currently stored in \texttt{edx}.
Finally, the \texttt{ret} instruction is encoded as three IL instructions:
\begin{enumerate}
	\item 
          \texttt{LDM	 R\_ESP:32, V\_01:32}: load the value stored
          in the memory address pointed to by \texttt{esp}, into the
          temporary register  \texttt{V\_01}. This address popped from 
          the stack is the return address of the function and from where
          the program should resume execution. 
	\item 
	\texttt{ADD	 R\_ESP:32, 4:32, R\_ESP:32}: pop from the
        stack, by incrementing the \texttt{esp} register value by $4$.  
	\item 
	\texttt{JCC	 1:1, V\_01:32}: resume execution at the instruction 
	immediately following the call within the calling function. Jump to 
	the return address.
\end{enumerate}

\begin{figure}[htbp]
	\centering
	\begin{adjustbox}{width=0.7\linewidth}
		\begin{minipage}{\linewidth}
			\captionsetup{justification=centering}
			\vspace{0.3em}
			\begin{lstlisting}[
				language={[x86masm]Assembler},
				basicstyle=\ttfamily\footnotesize,
				frame=single,
				columns=fullflexible,
				keepspaces=true,
				breaklines=false,
				breakatwhitespace=false,
				xleftmargin=0pt,
				framexleftmargin=0pt,
				tabsize=1]
				Asm:	mov eax, ecx				    IL:		STR R_ECX:32, R_EAX:32
				
				Asm:	add eax, edx				    IL:  	ADD R_EAX:32, R_EDX:32, R_EAX:32
				
				Asm:	ret         				    IL:  	LDM R_ESP:32, , V_01:32
			    						                      ADD R_ESP:32, 4:32, R_ESP:32
			    						                      JCC 1:1, , V_01:32
				
			\end{lstlisting}
		\end{minipage}
    \end{adjustbox}
    \caption{Assembly code (on the left), and its IL representation on
    the right.}
    \label{fig:exampleStep2}
\end{figure}

\mysubsection{Step 3: Converting Intermediate Language Statements to a Logical Formula.}
In the third step, we encode the IL instructions into a logical 
formula for the SMT solver, in SMT-LIB
format~\cite{CaFoTi16}. Because each IL instruction is standalone and atomic, it can
be translated independently of other instructions.
However, we still need to account for the chronological order of
instructions --- for example, the same register can be used by
different commands, and take different assignments at different points
in time during the execution. As an illustrative example, consider the
assembly code in Fig.~\ref{fig:exampleStep3Assembly}.  A simple
translation of the assembly code above will yield the SMT assertions
in Fig.~\ref{fig:exampleStep3SMT}(a).  Clearly, this formula is
unsatisfiable, while the assembly code is feasible. To circumvent this
problem, we applied an encoding that uses a fresh set of variables for
every time step of the execution; see
Fig.~\ref{fig:exampleStep3SMT}(b).

\begin{figure}[htbp]
	\centering
	\begin{adjustbox}{width=0.7\linewidth}
		\begin{minipage}{\linewidth}
			\captionsetup{justification=centering}
			\vspace{0.3em}
			\begin{lstlisting}[language={[x86masm]Assembler},basicstyle=\ttfamily\footnotesize,frame=single]
				Mov esi, 0x48
				mov esi, 0x2007
			\end{lstlisting}
		\end{minipage}
              \end{adjustbox}
              \caption{Assembly code.}
              \label{fig:exampleStep3Assembly}
\end{figure}

\begin{figure}
	\centering
	\begin{subfigure}[t]{0.48\textwidth}
		\centering
		\begin{adjustbox}{width=\linewidth}
			\begin{minipage}{\linewidth}
				\captionsetup{justification=centering}
				\vspace{0.3em}
				\begin{lstlisting}[style=asmstyle]
					; Variable declarations
					(declare-fun esi () Int)
					
					; Constraints
					(assert (= esi 0x48))
					(assert (= esi 0x2007))
				\end{lstlisting}
			\end{minipage}
                      \end{adjustbox}
                      \caption{A naive translation.}
	\end{subfigure}
	\hfill
	\begin{subfigure}[t]{0.48\textwidth}
		\centering
		\begin{adjustbox}{width=\linewidth}
			\begin{minipage}{\linewidth}
				\captionsetup{justification=centering}
				\vspace{0.3em}
				\begin{lstlisting}[style=asmstyle]
					; Variable declarations
					(declare-fun esi_step1 () Int)
					(declare-fun esi_step2 () Int)
					
					; Constraints
					(assert (= esi_step1 0x48))
					(assert (= esi_step2 0x2007))
				\end{lstlisting}
			\end{minipage}
                      \end{adjustbox}
                      \caption{A correct translation.}

                    \end{subfigure}
                    \caption{Translating IL code to SMT-LIB.}
                    \label{fig:exampleStep3SMT}
\end{figure}

Going back to our running example from Fig.~\ref{fig:exampleStep1},
the conversion of the IL instructions from Fig.~\ref{fig:exampleStep2}
yield the SMT-LIB assertions in
Fig.~\ref{fig:exampleStep3RunningExample}.  The first part of the
SMT-LIB encoding (lines 1--6) includes declaring the different
registers: \texttt{EAX, ECX, EDX, EIP}, and \texttt{ESP}. This is
followed by a declaration of a 32-bit word-addressable array
\texttt{MEM} (lines 8--9), used for encoding the loading of values
from memory; specifically, this definition implies that each memory
address is a 32-bit word, and that each memory entry is also a 32-bit
word. Thus, \texttt{MEM[addr]} returns the 32-bit word stored at
address \textit{addr}.  Then, in lines 11--16 we decelerate of the post-state
registers \texttt{EAX\_output, ECX\_output, EDX\_output, EIP\_output}
and \texttt{ESP\_output}, which represent the values these registers
hold after the code finishes running.

\begin{figure}[htb] 
	\centering
	\begin{minipage}[t]{0.48\textwidth}
		\begin{lstlisting}[style=libstyle]
			; Registers (32-bit)
			(declare-fun EAX () (_ BitVec 32))
			(declare-fun ECX () (_ BitVec 32))
			(declare-fun EDX () (_ BitVec 32))
			(declare-fun EIP () (_ BitVec 32)) ; instruction pointer
			(declare-fun ESP () (_ BitVec 32)) ; stack pointer
			
			; Memory: 32-bit word-addressable
			(declare-fun MEM () (Array (_ BitVec 32) (_ BitVec 32)))
			
			; Post-state (after executing instructions)
			(declare-fun EAX_output () (_ BitVec 32))
			(declare-fun ECX_output () (_ BitVec 32))
			(declare-fun EDX_output () (_ BitVec 32))
			(declare-fun EIP_output () (_ BitVec 32))
			(declare-fun ESP_output () (_ BitVec 32))
			
			; mov eax, ecx
			(define-fun step1_eax () (_ BitVec 32) ECX)
			;add eax, edx
			(define-fun step2_eax () (_ BitVec 32) (bvadd step1_eax EDX))
		\end{lstlisting}
	\end{minipage}
	\hfill
	\begin{minipage}[t]{0.48\textwidth}
		\begin{lstlisting}[style=libstyle, firstnumber=23]
			;ret
			; load return address  
			(define-fun V_01 () (_ BitVec 32) (select MEM ESP))
			; increment stack pointer
			(define-fun step1_esp () (_ BitVec 32) (bvadd ESP #x00000004))
			; continues code execution from the calling func
			(define-fun step2_eip () (_ BitVec 32) V_01)
			
			(assert (= EAX_output step2_eax))
			(assert (= ECX_output ECX)) ; unchanged
			(assert (= EDX_output EDX)) ; unchanged
			(assert (= ESP_output step1_esp))
			(assert (= EIP_output step2_eip))
			
			(check-sat)
			(get-model)
		\end{lstlisting}
	\end{minipage}
	\caption{The full SMT-LIB formula (broken into two boxes due
          to spacing constraints).}
	\label{fig:exampleStep3RunningExample}
\end{figure}

Next, in lines 18--21 we declare two new variables, representing 
\texttt{eax} after assigning the value in \texttt{ecx} and then 
again after adding to it the value of \texttt{edx}. 
The last three definitions, in lines 23--29, are derived from the 
\texttt{ret} instruction. The first statement assigns the value 
stored in memory where the \texttt{esp} register points to, to 
the variable \texttt{V\_01}. The second statement increments the 
value of the \texttt{esp} register, and  pops the value at the top 
of the stack (the return value). 
The third statement assigns the value stored in the temporary register 
\texttt{V\_01} to the \texttt{eip} register, and sets the execution 
to resume at the instruction immediately after the call, within the 
calling function.       
Finally, in lines 31--35, we have assertions that force the post-state 
variables to hold the values assigned to the corresponding registers 
at the end of the execution.

\mysubsection{Step 4: Solving the Logical Formula using an SMT Solver.}
In the fourth and final step, we employ an SMT solver to dispatch the 
resulting formula. The use of SMT-LIB format, which all modern solvers 
support, makes it straightforward to use off-the-shelf solvers, or even 
a portfolio thereof.

Returning to the encoding from Fig.~\ref{fig:exampleStep3RunningExample},
suppose our RE goal is to find an input for which the program outputs
the value 8. We add the assertions
\begin{align*}
  (\text{assert } (= \text{EAX\_output } \#0x00000008))
\end{align*}
to the encoding, and invoke the SMT solver. In this case, the solver 
will return SAT, as well as a satisfying assignment. One possible 
such assignment is $c=5$ and $d=3$, which indeed indicates an input 
for which the program produces the desired output.

Additional details about the ReSMT approach and pipeline
 appear in Sec~\ref{sec:case_study} and in 
Appendix~\ref{sec:deep_analysis}.

\section{ReSMT: Features and Design}
\label{sec:featuresAndDesign}

\subsection{Main Features}

We implemented the aforementioned technique in a new tool, named ReSMT. 
The tool, implemented in Pyhton, takes as input four parameters: 
\begin{inparaenum}[(i)]
	\item
	a path to the binary file to be analyzed;
	\item
	the memory address of the target function;
	\item
	the size in bytes of the target function; and
	\item 
	a path to a Json file containing the RE query to be solved.
\end{inparaenum}
An illustrative example and an explanation on the properties that can
be specified appears in
Appendix~\ref{ReSMT_tool}.

As output, the tool can either return UNSAT, indicating that no
solution was found for the RE query; or SAT, accompanied by a
satisfying assignment, which the engineer can then use to construct
the input of interest. A third option is that the tool will timeout
due to the complexity of the query.

A key feature of ReSMT is that it can solve RE queries automatically,
with no human-in-the-loop. As we later describe, this is often
achieved very efficiently. To the best of our knowledge, no other tool
offers a complete, end-to-end solution that achieves this level of
performance.

Another feature of ReSMT is that it is obfuscation-agnostic: whereas
many automated analysis tools are effective only against specific
obfuscators~\cite{Ha2019, Bl2021}, our tool is not tied to any
particular vendor or obfuscation product, and can handle a verity of
obfuscation techniques.

\subsection{Tool Design}

As discussed in Section~\ref{sec:our_approach}, ReSMT has three main
components:
\begin{inparaenum}[(i)]
	\item
    	a bytecode disassembler;
	\item
		a translation \& simulation framework, in order to translate the assembly code to 
		Intermediate Language (IL); and
	\item
		an SMT solver. 
\end{inparaenum}
In ReSMT, we used existing tools and packages as backends, making the
necessary adjustments and modifications. The design we used is
modular, so that additional backends and features can be integrated
into ReSMT in the future.

\mysubsection{Bytecode Disassembler.}  For bytecode disassembly, ReSMT
uses several standard, third-party packages: the python libraries
\textit{pefile}~\cite{pefile} and \textit{pybfd}~\cite{pybfd} for
loading \textit{ELF} and \textit{PE} files, and the
\textit{Capstone}~\cite{AnQuCapstoneCommunity2025} framework for
disassembly of binary files. The \textit{Capstone} dissembler
supports a wide variety of CPU architectures, while \textit{pybfd} and
\textit{pefile} support the parsing of \textit{PE} and \textit{ELF}
files, regardless of the CPU architecture these files were compiled
for. These libraries are often used as a pipeline feeding into
OpenREIL~\cite{DuPo}.

\mysubsection{Simulation Framework and Translator to IL.}  
For this part of the process, our tool uses the OpenREIL
framework~\cite{DuPo} as a backend, as well as its built-in output
format as the intermediate language. The framework supports three
different CPU architectures: x86, ARM and PowerPC, and each of these 
has a dedicated translator to IL instructions. 
For simplicity, in our initial
implementation we focused on a subset of just 17 ``core'' IL
instructions, which are sufficient for many tasks, and which can be
used to encode almost all of the assembly instructions in the x86 
architecture. In the future, we plan to extend the tool to support 
additional CPU architectures and instructions as needed.

In ReSMT, we used additional libraries to extend the functionality of
the OpenREIL emulator to support symbolic execution, so that it could
be used to execute IL instructions and emulate the execution of the
program's code, including its effect on the PC architecture: memory,
stack, heap, registers, etc. These extensions are not part of OpenREIL's 
core; but our efforts were inspired by previous attempts that we discovered
online~\cite{Ol}. The virtual machine uses a flat memory model, and 
supports an unlimited number of auxiliary variables.
More specifically, the additional library we used on top of the 
OpenREIL emulator affords the following functionality:
\begin{inparaenum}[(i)]
	\item
	storing symbolic values;
	\item
	propagating symbolic expressions through arithmetic;
	\item 
    translating the CPU state into SMT-LIB constraints; and
	\item 
    allowing REIL-based symbolic execution.
\end{inparaenum}
This adds the support and functionality needed for a basic symbolic 
execution of IL instructions, in a way that allows us to receive a set 
of instructions in IL and output their effects as logical 
assertions.

\mysubsection{The Z3 SMT solver.}  Our tool uses the Z3 SMT
solver~\cite{DeBj08} as a backend for dispatching the
results verification queries (Step 4 of the technique).  Z3 is a
state-of-the art solver, widely used in software analysis, hardware
verification and verification tools. It is open-source tool, accepts
input in VNNLIB format, and supports the theories of arithmetic,
bit-vectors, arrays, and combinations thereof. Still, the modular
design of ReSMT should allow to use other SMT solvers as backends, in
a straightforward manner.

\section{Case Study: Find the Key}
\label{sec:case_study}

\subsection{Setup}

\begin{wrapfigure}[9]{r}{0.5\textwidth}
  \vspace{-1.3cm}
	\begin{center}
		\includegraphics[width=\linewidth, keepaspectratio]{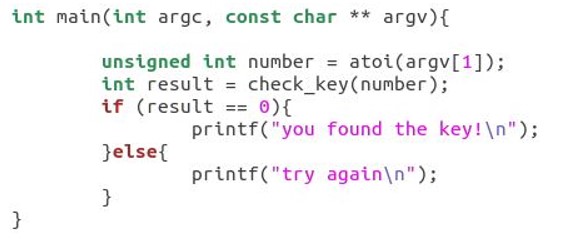}
	\end{center}
	\caption{Code for the main function of the program.}
	\label{fig:main_func_c_code}
\end{wrapfigure}

To evaluate ReSMT, we focused on a well-known problem in
cybersecurity: analyzing obfuscated code to extract from it hidden
information, such as an embedded key. We created several C programs
that share the same logic, but which significantly differ in the level
of obfuscation applied to them --- and consequently, in how difficult
it is to analyze them.

The first program in our suite is the unobfuscated baseline (see
Fig.~\ref{fig:main_func_c_code}): the program prompts for a password
(a 32-bit integer), and prints a success message if the supplied key
is correct, or a failure message otherwise. This piece of code is
identical across all the program variants that we used, and we assume
it is known to the user trying to reverse-engineer the program. The
source for the routine that validates the user-supplied key, namely
\texttt{check\_key()}, is provided to the user only in compiled,
binary form. The key value stored therein was chosen arbitrarily.
This function is where we applied the obfuscation techniques, 
creating five additional variants, each more complex than the last.
	
\mysubsection{The Obfuscated Programs.}  To obfuscate the
\texttt{check\_key()} function, we used a commercial tool called
\emph{PELOCK}~\cite{Wo2023}. This tool requires
a paid license, and offers a variety of obfuscation methods and
techniques specifically designed for assembly code. 

PELOCK supports a variety of obfuscation techniques, of which we
selected the following for our case-study:	
\begin{inparaenum}[(i)]
\item 
  changing the code execution flow to a nonlinear path;
\item 
  mutating original opcodes into series of other, equivalent instructions;
\item
  inserting ``fake'' instructions --- instructions that have no effect on 
  the outcome, because the control flow never reaches them (and so
  they are ``dead code'');
\item
  inserting fake jumps ---  two opposite conditional jump instructions are 
  placed one after the other with the same destination; and
\item
  inserting ``junk'' instructions between original instructions --- 
  instructions that do not affect the program state, but are simply
  there to mislead disassemblers.
\end{inparaenum}
We denote the baseline, unobfuscated program as $P_0$, and the
obfuscated variants as $P_1,\ldots,P_5$. Each program $P_i$ is constructed
incrementally: for every $i>1$,
$P_i$ is produced from $P_{i-1}$ by applying some or all of the
obfuscation techniques $1,\ldots,5$. As a result, every successive
variant introduces additional layers of obfuscation and complexity.
Program $P_5$ is therefore the most heavily obfuscated version, with
all techniques applied multiple times.

\mysubsection{$P_1$.} In $P_1$, the \texttt{check\_key()} function is
transformed by inserting jump instructions between original
operations. This alteration in control flow does not affect the semantics,
but increases visual clutter: the function now has \texttt{10}
instructions, and takes up \texttt{19} bytes comparing to \texttt{6} 
instructions and \texttt{11} bytes in the original, unobfuscated, 
program $P_0$. This transformation is fairly straightforward to undo;
see Fig.~\ref{fig:assembly_p1} in the appendix for a visualization.

\mysubsection{$P_2$ and $P_3$.}  $P_2$ and $P_3$ apply substantially
stronger, layered obfuscation. Both functions contain much additional
code: $P_2$ has \texttt{59} instructions and takes up \texttt{231}
bytes, whereas $P_3$ has \texttt{167} instructions and takes up
\texttt{953} bytes; see Fig.~\ref{obfuscated_program_2_3} in the
appendix. Both $P_2$ and $P_3$ were produced by applying the same
obfuscation techniques; but for $P_3$, the
number of obfuscation iterations was greater.

\mysubsection{$P_4$.} In $P_4$, the \texttt{check\_key()} function
has \texttt{3,219} instructions and takes up \texttt{18,200}
bytes. It is now quite hard to manually reverse-engineer; see
Fig.~\ref{obfuscated_program_4}. 
This version extends the previous program ($P_3$) and applies the same 
obfuscation techniques, but at a much higher intensity and complexity
level. 

\mysubsection{$P_5$.}  In $P_5$, the function has \texttt{7,680}
instructions and \texttt{24,156} bytes, and is extremely difficult to
manually analyze. The control flow is heavily broken by many orphan
jumps, and the state-of-the-art IDA Pro deobfuscator cannot produce a
coherent CFG due to numerous disconnected blocks and invalid jump
targets. See Fig.~\ref{obfuscated_program_5}; as well as
Appendix~\ref{Case_Study_appendix} for additional information
regarding the obfuscated programs $P_1$--$P_5$.

\mysubsection{Reverse Engineering Goal.}
The RE goal is to find the hard-coded hidden key in the function's 
code, in order to get the program to print the congratulatory message
to the screen.

\subsection{Reverse-Engineering $P_0,\ldots,P_6$}
\label{sec:programs_analyzing}
\begin{wrapfigure}[12]{r}{0.5\textwidth}
  \vspace{-0.9cm}
	\begin{center}
		\includegraphics[width=\linewidth, keepaspectratio]{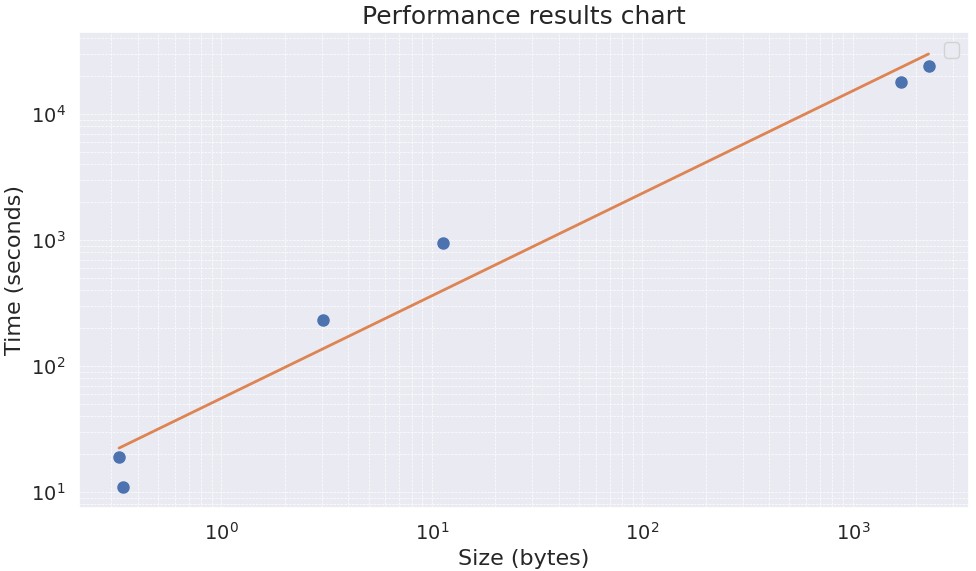}
	\end{center}
	\caption{The runtime of ReSMT as a function of code size.}
	\label{fig:evaluation}
\end{wrapfigure}
\sloppy
We now describe the reverse-engineering process of each of the
programs, $P_0,\ldots,P_5$. Since we are not aware of any tools with
similar functionality to that of ReSMT, we compared its performance to
that of two baselines: a reverse-engineering attempt by the automated
tool GPT-5.1 (the latest model currently available)~\cite{chatGPT5.1}, 
and to a manual reverse-engineering attempt by an RE expert with 
approximately 4 years of experience.
The results are summarized in Fig.~\ref{fig:evaluation} and in
Table~\ref{performance_table}. In addition to ReSMT's total runtime,
we also provide a breakdown into each of the three major steps:
instruction extraction and disassembly (the Inst.~Extraction column),
translation and emulation (the Trans.~\&~Emu. column), and
solving. All automated experiments were conducted on a virtual machine
with an Intel i3-4310U (2.0--2.6\,GHz) and 4\,GB RAM. We note that
using this setup, conducting a brute-force search to sample all
possible $2^{32}$ keys did not terminate in 48 hours.

\begin{table}[h!]
	\centering
	
	\renewcommand{\arraystretch}{1.2}
	\setlength{\tabcolsep}{2pt}
	
	\begin{tabular}{|
			>{\centering\arraybackslash}m{0.05\linewidth}|
			>{\centering\arraybackslash}m{0.11\linewidth}|
			>{\centering\arraybackslash}m{0.15\linewidth}|
			>{\centering\arraybackslash}m{0.11\linewidth}|
			>{\centering\arraybackslash}m{0.15\linewidth}|
			>{\centering\arraybackslash}m{0.11\linewidth}|
			>{\centering\arraybackslash}m{0.11\linewidth}|
			>{\centering\arraybackslash}m{0.11\linewidth}|
		}
		\hline
		\textbf{ } &
		\textbf{ Size (bytes)} &
		\textbf{ Inst. Extraction} &
		\textbf{ Trans. \& Emu.} &
		\textbf{Solving} &
		\textbf{Total Time} &
		\textbf{ ChatGPT 5.1} &
		\textbf{ Est. Manual Analysis Time} \\
		\hline
		
		\textbf{P$_0$} &
		11 &
		0.010453 s &
		0.046039 s &
		0.222008 s &
		0.34215 s &
		seconds &
		seconds \\
		\hline
		
		\textbf{P$_1$} &
		19 &
		0.000978 s &
		0.07581 s &
		0.179239 s &
		0.26518 s &
		seconds &
		seconds \\
		\hline
		
		\textbf{P$_2$} &
		231 &
		0.00657 s &
		1.21698 s &
		1.167089 s &
		3.05031 s &
		$\sim$1 min &
		about an hour \\
		\hline
		
		\textbf{P$_3$} &
		953 &
		0.011279 s &
		7.521078 s &
		2.126926 s &
		11.3423 s &
		1--2 min &
		few hours \\
		\hline
		
		\textbf{P$_4$} &
		18{,}200  &
		0.138933 s &
		26:48.73 min &
		52.819069 s &
		28:20.87 min &
		failed &
		few hours \\
		\hline
		
		\textbf{P$_5$} &
		24{,}156  &
		1.254092 s &
		37:27.68 min &
		40.286055 s &
		38:09.78 min &
		failed &
		hours to a day \\
		\hline
		
	\end{tabular}
	\vspace{0.3cm}
	\caption{Analyzing $P_0,\ldots,P_6$ with ReSMT, GPT, and the expected time for manual analysis.}
	\label{performance_table}
\end{table}

\vspace{-2.5em}

\begin{figure}[!htbp]
	\centering
	\begin{subfigure}[t]{0.45\textwidth}
		\includegraphics[width=\linewidth, keepaspectratio]{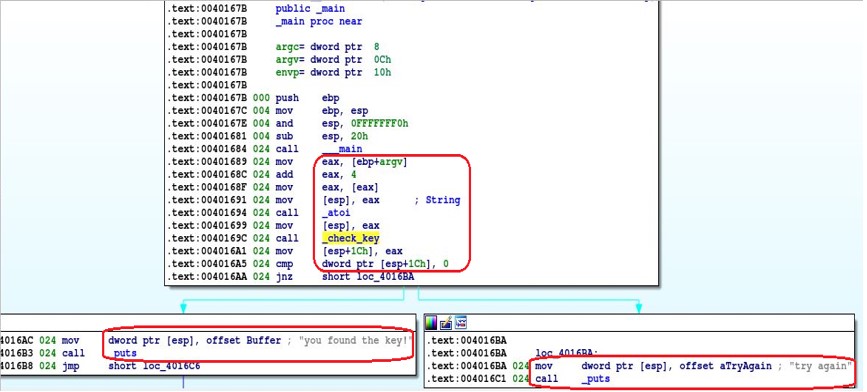}
		\label{fig:main_func}
	\end{subfigure}
	\hfill
	\begin{subfigure}[t]{0.45\textwidth}
		\includegraphics[width=\linewidth, keepaspectratio]{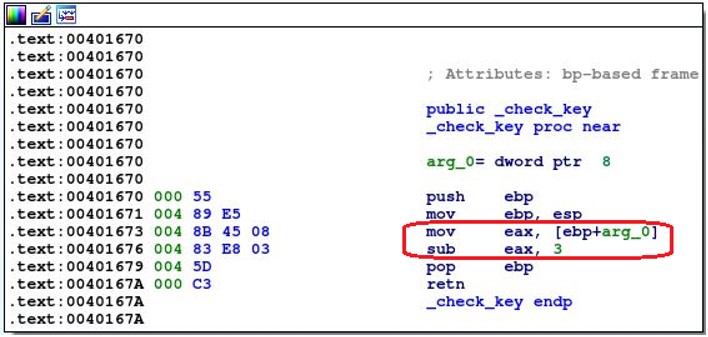}
		\caption{``check\_key'' function --- no obfuscation}
		\label{fig:check_key}
	\end{subfigure}
	\caption{Assembly code for the ``Main'' function
            of $P_0$ (on the left), and for the ``check\_key'' function
            (on the right).}
	\label{fig:original_program_assembly}
\end{figure}

\mysubsection{$P_0$.}  The original program, $P_0$, is fairly
straightforward to analyze manually. Static analysis using IDA
Pro~\cite{HexRays2024} (see Fig.~\ref{fig:original_program_assembly})
reveals that the \texttt{main()} function passes the user input to the
\texttt{atoi(string s)} function, then passes the resulting integer to
\texttt{check\_key(int key)}, and that the latter determines whether
the operation succeeds or fails.  Further inspection shows that the
integer key is transferred through the \texttt{EAX} register and
pushed onto the stack (\texttt{ESP}) after conversion by
\texttt{atoi}, indicating that the key is represented as a 32-bit
integer. The return value of \texttt{check\_key()} is also stored in
\texttt{EAX} and then moved into the memory location
\texttt{[ESP+1Ch]}, where it is compared against \texttt{0}. A zero
result triggers the success message; otherwise, the program outputs a
failure notification.  Examining the implementation of
\texttt{check\_key()} reveals that it subtracts the input key from the
constant value \texttt{0x03}, stores the result in \texttt{EAX}, and
returns it. Therefore, the correct key is \texttt{3}. Manually
reversing this non-obfuscated function by an experienced security
researcher takes a couple of minutes. Running this function through
ReSMT took about 350 milliseconds until the correct key was found.

\mysubsection{$P_1$--$P_5$.} The obfuscated programs, $P_1$ through
$P_5$, proved much more difficult to analyze manually. $P_1$ is poorly
obfuscated, and can easily be restored to its original form by
removing the jump statements --- and so manually reversing it takes
only a few minutes more than reversing the original program $P_0$. In
programs $P_2$--$P_5$, however, the complexity of the obfuscation
increases significantly. In order to reverse-engineer them, a
researcher first must determine which obfuscation techniques were
applied, and then seek relevant tools for deobfuscating them; or,
alternatively, write code for removing this obfuscation themselves. A
manual reversing attempt for programs $P_2$--$P_4$ would take between
an hour to several hours; whereas program $P_5$ is so complex that the effort 
might take up to a day.
For example, program $P_5$ is so cluttered
that even the IDA decompiler failed to build the CFG for the function, 
and could do no more than present the assembly code itself.
In the case of $P_4$, although IDA was able to present a CFG, it is
extremely cluttered, and is of little use.
See Sec.~\ref{Case_Study_appendix} of the appendix for additional details.

In sharp contrast, ReSMT succeeded in reverse-engineering each of the
programs $P_1$--$P_5$. For $P_1$--$P_3$, this took a few seconds;
whereas $P_4$ and $P_5$ took 28 minutes and 38 minutes, respectively.
A more thorough description of ReSMT's operation on these programs appears
in Appendix~\ref{sec:deep_analysis}.

\mysubsection{Discussion.}  Inspecting the results in
Fig.~\ref{fig:evaluation}, we see the ReSMT scales quite well --- and
succeeds in tackling even $P_5$ in under 40 minutes, despite its
significant size and complexity. We find that these results showcase
the potential of our approach. We further observe that the runtime of
the ``instructions extraction'' step grows linearly in the size of the
code, but remains negligible compared to the other steps. The runtime
required for the ``Translation and Emulation'' step grows
linearly in the size of the code, and takes roughly 95\% of the
total runtime. Thus, this step is the bottleneck when dealing with
larger programs, and we plan to optimize this step in future
work. Finally, the ``Solving'' step increases linearly with the size
of the obfuscated program, and is again negligible. Comparing ReSMT to 
ChatGPT 5.1 and manual analysis, we find that ReSMT clearly outperforms 
both. ChatGPT 5.1 solves the first four programs but fails on $P_4$ and $P_5$.

\vspace{-1.0em}
\section{Related Work}
\label{sec:Related_Work}
\vspace{-0.5em}

The use of SMT solvers in various code analysis
tasks~\cite{FiSlMiMi2016}, and in particular for code deobfuscation, is
quite common. We mention here a few notable approaches.

Goomba~\cite{Pe2023} for IDA Pro~\cite{HexRays2024} matches MBA-based 
obfuscated code to known patterns and uses Z3 to verify their equivalence.  
GhiHorn \cite{Ge2021} is a plugin for the Ghidra reverse-engineering
framework \cite{Ghidra}, designed to support path analysis and discovery
in both assembly and high-level code. It leverages an SMT 
solver to reason about program conditions and reconstruct feasible 
execution paths. Although GhiHorn provides basic memory simulation, its
model is more limited compared to our solution, which emulates multiple
memory domains such as the stack, heap, and other regions.
		
Opticode \cite{An2013} is a framework for deobfuscating obfuscated machine
code using two methods:
\begin{inparaenum}[(i)]
	\item
	Code optimization using the LLVM IR intermediate language, in
  	order to remove garbage code; and
	\item
	Removal of dead code (opaque predicates) using an SMT solver. 
\end{inparaenum}
In contrast, our approach can tackle complex RE goals. 
	
The BINSEC platform \cite{DaBaTaMoFePoMa16} analyzes obfuscated malware by 
combining dynamic disassembly with both static and dynamic program 
analysis. It leverages SMT solving to eliminate opaque predicates and 
remove dead code. It would be interesting to explore the synergies 
between ReSMT and BINSEC ---  perhaps by pipelining the two, first running  
BINSEC  to simplify the code before feeding it to ReSMT.

SATURN~\cite{GaFa2019} is a deobfuscation framework that lifts obfuscated 
binaries into LLVM IR for advanced analysis and simplification. It uses 
SMT solving to defeat various obfuscation techniques. Unlike ReSMT, it 
focuses on static analysis and reconstructs code in a form close to the 
original.


\section{Conclusion \& Future Work}
\label{sec:Discussion}
\vspace{-0.5em}

We introduced ReSMT, a tool for automated reverse engineering of 
obfuscated code. In our case study, ReSMT successfully handled various 
PELOCK~\cite{Wo2023} obfuscation techniques and automatically recovered 
the hidden key, even when manual analysis was difficult. These results 
demonstrate ReSMT's potential for real-world applications, such as 
analyzing obfuscated malware.

Our work can be extended in two main directions. First, ReSMT could be 
applied to additional obfuscation techniques to identify where it succeeds 
and where it struggles. Challenging cases might be addressed by trying 
other SMT solvers as backends or by using more advanced SMT-LIB encodings.

A second direction for future work is to study the scalability of the 
tool on more complex inputs. Although our case study suggests that ReSMT's 
runtime grows roughly linearly with the size of the obfuscated code, this 
trend may be affected by hardware and program complexity. To improve 
scalability, we plan to introduce caching, storing frequently encountered 
IL instructions and their encodings to reduce redundant SMT-LIB generation.


\bibliographystyle{abbrv}
\bibliography{Biblography_arxiv}

@misc{Wo2023, 
	title        = {{PELock Obfuscator}},
	author       = {Wojcik, B.},
	year         = {2023},
	note         = {\url{https://www.pelock.com/products/obfuscator}}
}

@misc{guardsquare,
        Author = {GuardSquare},
	title        = {{ProGuard 7.5}},
	note 		 = {\url{https://www.guardsquare.com/proguard}},
	year         = {2024}
}

@misc{jscrambler,
	title        = {{Jscrambler}},
	Author = {Jscrambler},
	note         = {\url{https://jscrambler.com/}},
	year         = {2024}
}

@misc{preemptive2024a,
	title        = {{Dotfuscator}},
	year         = {2024},
	note         = {\url{https://www.preemptive.com/products/dotfuscator/}},
	author = {PreEmptive}
}

@misc{VMP,
        author = {VMProtect},
	title        = {{VMProtect}},
	year         = {2024},
	note         = {\url{https://vmpsoft.com/}}
}

@misc{TM,
        author = {Oreans},
	title        = {{Themida}},
	year         = {2024},
	note         = {\url{https://www.oreans.com/Themida.php}}
}

@misc{Lu2024, 
	title        = {{Obfuscation in Software: Definition and Techniques}},
	author       = {Awati, R. and Lutkevich, B.},
	note	    = {Technical Report. \url{https://www.techtarget.com/searchsecurity/definition/obfuscation#:~:text=Deobfuscation%20techniques%20can%20be%20used,particular%20point%20in%20the%20program.}},
	year         = {2024}
}

@misc{Ba2025,
	author       = {{Basatwar, G.}},
	title        = {{Code Obfuscation}},
	note		 = {Technical Report. \url{https://www.appsealing.com/code-obfuscation/}},
	year		 = {2025}
}

@inproceedings{DaBaTaMoFePoMa16,
	title 		 = {{BINSEC/SE: A Dynamic Symbolic Execution Toolkit for Binary-Level Analysis}},
	author 		 = {David, R. and Bardin, S. and Ta, T. D. and Mounier, L. and Feist, J. and Potet, M. L. and Marion, J. Y.},
	booktitle 	 = {Proc. 23rd Int. Conf. on Software Analysis, Evolution, and Reengineering (SANER)},
	year      	 = {2016}
}

@misc{HexRays2024,
        author = {Hex-Rays},
	title        = {{IDA Pro: Interactive Disassembler and Debugger for Binary Analysis}},
	year         = {2025},
	note         = {\url{https://hex-rays.com/ida-pro/}}
}

@misc{Pe2023,
	title 		 = {{Binary Deobfuscation with gooMBA}},
	author 		 = {Petrov, A.},
	note 		 = {Technical Report. \url{https://hex-rays.com/blog/deobfuscation-with-goomba/}},
	year 		 = {2023}
}

@misc{Ge2021,
	title 		 = {{GidHorn Path Analysis}},
	author 		 = {Gennari, J.},
	note		 = {Technical Report. \url{https://insights.sei.cmu.edu/blog/ghihorn-path-analysis-in-ghidra-using-smt-solvers/}},
	year 		 = {2021}
}

@misc{An2013,
	Title 		 = {{OptiCode: Machine Code Deobfuscation for Malware Analysis}},
	Author 		 = {Anh Quynh, N.},
	Note 		 = {Technical Report. \url{https://www.data.proidea.org.pl/confidence/11edycja/NGUYEN_Anh_Quynh.pdf}},
	Year 		 = {2013}
}

@misc{Ga20,
	title     	 = {{Code Deobfuscation by Program Synthesis-Aided Simplification of Mixed Boolean-Arithmetic Expressions}},
	author    	 = {G\`aomez i Montolio, A.},
	Note = {\url{https://arnaugamez.com/assets/theses/bsc.pdf}},
	year		 = {2020}
}

@article{LyMiZhXu2020,
	title     	 = {{Layered Obfuscation: a Taxonomy of Software Obfuscation Techniques for Layered Security}},
	author    	 = {Xu, H. and Zhou, Y. and Ming, J. and Lyu, M.},
        journal = {Cybersecurity},
	volume = {3},
	number = {1},
	year		 = {2020}
}

@misc{Ghidra,
        author = {{National Security Agency}},
	title        = {{Ghidra: Software Reverse Engineering Framework}},
	note         = {\url{https://github.com/NationalSecurityAgency/ghidra}},
	year         = {2025}
}

@misc{chatGPT5.1,
author = {OpenAI},
	title        = {{ChatGPT 5.1}},
	note 		 = {\url{https://openai.com/index/gpt-5-1/}},
	year         = {2025}
}

@MISC{CaFoTi16,
  author =	 {Barrett, C. and Fontaine, P. and Tinelli, C.},
  title =	 {{The Satisfiability Modulo Theories Library (SMT-LIB)}},
  howpublished = {\url{www.SMT-LIB.org}},
  year =	 {2016},
}

@misc{DuPo,
	title	 	= {{REIL: A Platform-Independent Intermediate Representation of Disassembled Code for Static Code Analysis}},
	author	 	= {Dullien, T. and Porst, S.},
	note		= {\url{https://static.googleusercontent.com/media/www.zynamics.com/en//downloads/csw09.pdf}},
	year = {2009},
}

@misc{Jj2025,
	title        = {{Deobfuscation Techniques: Peephole Deobfuscation}},
	author       = {Jedynak, J.},
	note		 = {Technical report. \url{https://cert.pl/en/posts/2025/04/peephole-deobfuscation/}},
	year         = {2025}
}

@misc{Bh2024,
	title        = {{Generative AI in Reverse Engineering Obfuscated Code}},
	author       = {Bhairav, S.},
	note		 = {Technical report. \url{https://www.metriccoders.com/post/generative-ai-in-reverse-engineering-obfuscated-code}},
	year         = {2024}
}

@misc{Ha2019,
	title        = {{Defeating Compiler-Level Obfuscations Used in APT10 Malware}},
	author       = {Haruyama, T.},
	note         = {Technical report. \url{https://blogs.vmware.com/security/2019/02/defeating-compiler-level-obfuscations-used-in-apt10-malware.html}},
	year         = {2019}
}

@misc{Bl2021,
	title        = {{Automated Detection of Obfuscated Code}},
	author       = {Blazytko, T.},
	note         = {Technical report. \url{https://www.emproof.com/automated-detection-of-obfuscated-code/?utm_source=chatgpt.com}},
	year         = {2021}
}

@misc{AnQuCapstoneCommunity2025,
	title        = {{Capstone: a Lightweight Multi-Architecture Disassembly Framework}},
	author       = {Anh Quynh, N.},
	note         = {\url{https://github.com/capstone-engine/capstone}},
	year         = {2025}
}

@misc{pybfd,
	title        = {{PyBFD: A Python Interface to the GNU Binary File Descriptor (BFD) Library}},
	author = {Ortega, A.},
	note         = {\url{https://github.com/Groundworkstech/pybfd}},
	year         = {2013}
}

@misc{pefile,
	title        = {{PEFfile: a Python Module to Read and Work with PE (Portable Executable) Files}},
	author       = {Carrera, E.},
	note         = {\url{https://github.com/erocarrera/pefile}},
	year         = {2024}  
}

@misc{objdump,
        author = {{Free Software Foundation}},
	title        = {{ObjDump --- Display Information from Object Files}},
	note         = {\url{https://man7.org/linux/man-pages/man1/objdump.1.html}},
	year         = {2025}
}

@misc{windbg,
author = {Microsoft},
	title        = {{WinDBG --- Windows Operating System Debugger}},
	note         = {\url{https://learn.microsoft.com/en-us/windows-hardware/drivers/debugger/}},
	year         = {2025}
}

@misc{Ol,
	title		= {{Toy Project}},
	author		= {Oleksiuk, D.},
	note		= {Github reposetory. \url{https://github.com/Cr4sh/openreil/blob/master/tests/test_kao.py}}
}

@Book{Ei05,
  author = {Eilam, E.},
  title = {Reversing: Secrets of Reverse Engineering},
  Publisher = {Wiley Publishing},
  Year = {2025},
  }

@inproceedings{DeBj08,
  Title = {{Z3: An Efficient SMT Solver}},
  Author = {De Moura, L. and Bjorner, N.},
  Booktitle = {Proc. 14th Int. Conf. on Tools and Algorithms for the Construction and Analysis of Systems (TACAS)},
  pages = {337--340},
  Year = {2008}
}

@article{BeLa15,
 author = {Behera, C. and Lalitha Bhaskari, D.},
 title = {{Different Obfuscation Techniques for Code Protection}},
 journal = {Procedia Computer Science},
 volume = {70},
 pages = {757--763},
 year = {2015},
}

@incollection{BaSeSeTi09,
author = {Barrett, C. and Sebastiani, R. and Seshia, S. and Tinelli, C.},
title = {{Satisfiability Modulo Theories}},
booktitle = {Handbook of Satisfiability},
year = {2009},
publisher = {IOS Press}
}

@misc{GaFa2019,
	title   = {{SATURN --- Software Deobfuscation Framework Based on LLVM}},
	author  = {Garba, P. and Favaro, M.},
	note         = {Technical report. \url{https://arxiv.org/abs/1909.01752}},
	year         = {2019}
}

@inproceedings{YoYi2010,
	title     = {{Malware Obfuscation Techniques: A Brief Survey}},
	author    = {You, I. and Yim, K.},
	booktitle = {Proc. 5th Int. Conf. on Broadband, Wireless Computing, Communication and Applications (BWCCA)},
	year      = {2010},
}

@article{ScKaKiMeWe2016,
	title        = {{Protecting Software through Obfuscation: Can It Keep Pace with Progress in Code Analysis?}},
	author       = {Schrittwieser, S. and Katzenbeisser, S. and Kinder, J. and Merzdovnik, G. and Weippl, E.},
	journal      = {ACM Computing Surveys},
	year         = {2016},
}

@article{KuLa2015,
	title     = {{Different Obfuscation Techniques for Code Protection}},
	author    = {Kumar Behera, C.  and Lalitha Bhaskari, D.},
	journal   = {Procedia Computer Science},
	year      = {2015},
}

@article{FiSlMiMi2016,
	title     = {{Applying SMT Algorithms to Code Analysis}},
	author    = {Filipovic, M. and Sladic, G. and Milosavljevic, B. and Milosavljevic, G.},
	journal   = {Proc. 39th Int. Convention on Information and Communication Technology, Electronics and Microelectronics (MIPRO)},
	year      = {2016},
}

@misc{Ne1996,
	title   = {{A Survey of Reverse Engineering and Program Comprehension}},
	author  = {Nelson, M.},
	year    = {1996},
	note         = {Technical report. \url{https://arxiv.org/abs/cs/0503068}},
}

\clearpage
\appendix


\section{Applying the ReSMT Pipeline: A Thorough Description}
\label{sec:deep_analysis}

In this section, we more thoroughly describe ReSMT's solving process
when applied to our case study.  For clarity, we focus on the
\texttt{check\_key(int key)} function of the original program, which
is small and contains only a few instructions, making it easy to
examine manually and explain. We also consider the classical naive
approach ---a brute-force key search --- as a baseline.
 
\subsection{The Naive solution: Brute Force}

A naive approach for solving the challenge posed by our case-study is
to apply brute force, i.e., to try all possible keys 
until the correct one is found. The key is a signed 32-bit integer, 
yielding $2^{32} = 4,294,967,296$ possible values. Exhaustively trying 
every option is impractical: in our tests, a brute-force attempt ran for 
more than two days without completing (the ordering of the attempted
values was determined uniformly at random).

\subsection{Solving with ReSMT}

Following our approach, we construct a set of assertions and constraints 
in the theories of bit-vectors, linear arithmetic and arrays. These assertions represent 
the behavior of the \texttt{check\_key(int key)} function and encode the 
process by which the correct key is computed, expressed as a logical formula.

This formula is generated directly from the obfuscated assembly code by 
translating it into IL instructions and by incorporating the RE query 
provided by the analyst. Solving the resulting logical formula --- when a 
solution exists --- yields the correct key. This is because the formula encodes 
both the program’s key-generation logic and the relevant machine state 
(memory, stack, registers, etc.), along with the condition that identifies 
the correct key.

\begin{figure}[!htbp]
	\centering
	\includegraphics[width=0.8\textwidth, keepaspectratio]{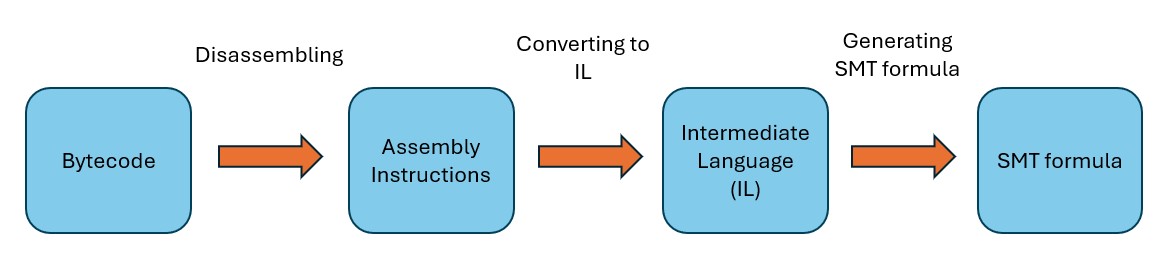}
	\caption{The high-level ReSMT pipeline.}
	\label{fig:Process_Flow}
\end{figure}

The model goes sequentially through the following four steps in order 
to deobfuscate the code and find the right key (see also Fig.~\ref{fig:Process_Flow}): 

\begin{enumerate}
	\item 
	Reading the bytecodes of the function \texttt{check\_key(int key)} from 
	the program and disassembling it into assembly code.
	\item 
	Converting the code to an intermediate language (IL) where each assembly 
	instruction is represented by one or more IL instructions.
	\item
	Generating from the IL instruction a logical formula consisting a set of 
	equations and constraints over the bit-vector, linear arithmetic and 
	array theories.
	\item
	 Solving the equation using an SMT solver.
\end{enumerate}

If the SMT solver succeeds in finding an satisfiable solution, then the correct 
key for the obfuscated assembly code has been recovered; the solution to 
the equation and constraint system corresponds directly to the correct key. 
If the solver returns UNSAT, 
 no solution exists.  A third possibility is that the tool times out due 
to query complexity, although this did not occur in our experiments.

We start by analyzing and explaining how our model handles the function 
\texttt{check\_key(int key)} of the original program ($P_0$):

\mysubsection{Bytecode disassembly.} The first step is extracting the 
function bytecode and disassembling it. 
The function consists of 11 bytes total, depicted in Fig.~\ref{fig:checkKeyBytecode}.

\vspace{1.5em}

\begin{figure}[htbp]
	\centering
	\begin{adjustbox}{width=0.7\linewidth}
		\begin{minipage}{\linewidth}
			\vspace{0.3em}
			\begin{lstlisting}[language={[x86masm]Assembler},basicstyle=\ttfamily\footnotesize,frame=single]
				55 89 E5 8B 45 08 83 E8 03 5D C3
			\end{lstlisting}
		\end{minipage}
              \end{adjustbox}
              \caption{The check\_key bytecode.}
              \label{fig:checkKeyBytecode}
\end{figure}

After extracting the bytecode of the \texttt{check\_key(int key)} function 
and disassembling it into assembly, we observe that the function takes a 
single user-provided argument and consists of six assembly
instructions, as depicted in Fig.~\ref{fig:checkKeyAssembly}.

\begin{figure}[htbp]
	\centering
	\begin{adjustbox}{width=0.7\linewidth}
		\begin{minipage}{\linewidth}
			\vspace{0.3em}
			\begin{lstlisting}[language={[x86masm]Assembler},basicstyle=\ttfamily\footnotesize,frame=single]
				public _check_key
				_check_key proc_near
				
				arg_0 = dword ptr 8
				
				000		push ebp
				001		mov ebp, esp
				003		mov eax, [ebp+arg_0]
				006		sub eax, 3
				009		pop ebp
				00A		retn
				
				_check_key endp
			\end{lstlisting}
		\end{minipage}
              \end{adjustbox}
              \caption{Basic bytecode and its corresponding assembly
                instructions.}
              \label{fig:checkKeyAssembly}
\end{figure}

The first two instructions form the function prologue, and the last two 
form the epilogue. These instructions set up and restore the stack frame 
to ensure correct execution of \texttt{check\_key(int key)} and seamless 
continuation of the program afterward. 

The third instruction, ``\texttt{mov eax, [ebp+arg\_0]}'', loads the user-provided key into the \texttt{EAX} register. The fourth instruction, 
\texttt{sub eax, 3}, subtracts the constant \texttt{3} from this value 
and stores the result back into \texttt{EAX}, which also holds the 
function’s return value.

\mysubsection{Translating Assembly code to IR.} In the second step we 
convert the assembly instructions of the function to an intermediate 
language (IL) where each IL instruction can be presented by an assertion 
or a term for the SMT solver. 
Each IL instruction is derived from a single assembly instruction, and 
one or more IL instructions can be derived from a single assembly instruction. 
Each set is constructed independently of the sets corresponding to other 
assembly-derived IL instructions.
In Fig.~\ref{fig:IL_translation} we can see the translation of the assembly code of the 
function \texttt{check\_key(int key)}.

\begin{figure}[htbp]
	\centering
	\begin{adjustbox}{width=0.7\linewidth}
		\begin{minipage}{\linewidth}
			\vspace{0.3em}
			\begin{lstlisting}[language={[x86masm]Assembler},basicstyle=\ttfamily\footnotesize,frame=single]
				Asm:	Push ebp
				
				IL:		STR		R_EBP:32, V_00:32
				STR		R_ESP:32, V_01:32
				SUB		V_01:32, 4:32, V_02:32
				STR		V_02:32, R_ESP:32
				STM		V_00:32, V_02:32
				
				
				Asm:	mov ebp, esp
				
				IL:		STR		R_ESP:32, V_00:32
				STR		V_00:32, R_EBP:32
				
				
				Asm:	eax, dword ptr [ebp + 8]
				
				IL:		STR		R_EBP:32, V_00:32
				ADD		V_00:32, 8:32, V_01:32
				LDM		V_01:32, V_02:32
				STR		V_02:32, R_EAX:32
				
				
				Asm:  	sub, eax, 3
				
				IL:		STR		R_EAX:32, V_00:32
				SUB		V_00:32, 3:32, V_01:32
				SUB		V_00:32, 3:32, V_02:32
				LT		V_00:32, 3:32, R_CF:1
				AND		V_02:32, FF:32, V_04:32
				OR		V_04:32, 0:32, V_03:8
				SHR		V_03:8, 7:8, V_05:8
				SHR		V_03:8, 6:8, V_06:8
				XOR		V_05:8, V_06:8, V_07:8
				SHR		V_03:8, 5:8, V_08:8
				SHR		V_03:8, 4:8, V_09:8
				XOR		V_08:8, V_09:8, V_10:8
				XOR		V_07:8, V_10:8, V_11:8
				SHR		V_03:8, 3:8, V_12:8
				SHR		V_03:8, 2:8, V_13:8
				XOR		V_12:8, V_13:8, V_14:8
				SHR		V_03:8, 1:8, V_15:8
				XOR		V_15:8, V_03:8, V_16:8
				XOR		V_14:8, V_16:8, V_17:8
				XOR		V_11:8, V_17:8, V_18:8
				AND		V_18:8, 1:8, V_20:8
				OR		V_20:8, 0:8, V_19:1
				NOT		V_19:1, R_PF:1
				XOR		V_00:32, 3:32, V_21:32
				XOR		V_02:32, V_21:32, V_22:32
				AND		10:32, V_22:32, V_23:32
				EQ		1:32, V_23:32, R_AF:1
				EQ		V_02:32, 0:32, R_ZF:1
				SHR		V_02:32, 1F:32, V_24:32
				AND		1:32, V_24:32, V_25:32
				EQ		1:32, V_25:32, R_SF:1
				XOR		V_00:32, 3:32, V_26:32
				XOR		V_00:32, V_02:32, V_27:32
				AND		V_26:32, V_27:32, V_28:32
				SHR		V_28:32, 1F:32, V_29:32
				EQ		1:32, V_30:32, R_OF:1
				STR		V_01:32, R_EAX:32
				
				
				Asm:  	pop ebp
				
				IL:		STR		R_ESP:32, V_00:32
				LDM		V_00:32, V_01:32
				ADD		V_00:32, 4:32, V_02:32
				STR		V_02:32, R_ESP:32
				STR		V_01:32, R_EBP:32
			\end{lstlisting}

		\end{minipage}
              \end{adjustbox}
              \caption{The assembly instructions and their
                corresponding IL instructions.}
              \label{fig:IL_translation}
\end{figure}

Each IL instruction is an atomic operation and each assembly instruction 
is represented by several IL instructions; and so the five assembly 
instructions are represented by dozens of IL instructions in total. 

We now examine the first assembly instruction and its translation to
IL. 
The first assembly instruction is --- \texttt{push ebp}. This instruction 
subtracts the value \texttt{4} from the \texttt{esp} register, which 
points to the top of the stack. Then, it stores the value contained 
in the \texttt{ebp} register at the top of the stack. So, the first 
IL instruction --- \texttt{STR  R\_EBP:32, V\_00:32} --- stores the value 
of the register \texttt{ebp} in the variable \texttt{V\_00} (these two 
variables are 32 bits). 
The next instruction --- \texttt{STR  R\_ESP:32, V\_01:32} --- stores the 
value of the register \texttt{esp} in the variable \texttt{V\_01}. 
The third instruction --- \texttt{SUB	V\_01:32, 4:32, V\_02:32} --- subtracts 
the value 4 from \texttt{V\_01} (this variable stores the value of 
\texttt{esp} register) and stores the result in the variable \texttt{V\_02}. 
The fourth IL instruction stores back in \texttt{esp} the value of 
 variable \texttt{V\_02}. 
In fact, the second through fourth IL instructions represents the 
subtraction of the value \texttt{4} from the \texttt{esp} register. 
The fifth IL instruction --- \texttt{STM   V\_00:32, V\_02:32} --- stores in 
memory the value of the variable \texttt{V\_00} (\texttt{ebp} register) 
in the memory where variable \texttt{V\_02} points to, which is the top 
of the stack (\texttt{esp} register).

\mysubsection{Converting IL Instructions to a Logical Formula.} In the 
third step of the model, the IL instruction sequence is translated 
into a unified logical formula for the SMT solver, which enables it to search 
for a satisfiable assignment. Each atomic IL instruction is mapped 
to its corresponding equation or constraint in the SMT formula, thereby encoding the 
semantics of the original assembly operations. After this translation is 
completed, we introduce a final constraint, \texttt{EAX == 0}, which brings 
the formula to its complete form. This condition ensures that any satisfying 
assignment represents the correct key, that is, the input that causes 
\texttt{check\_key(int key)} to return zero. In this way, the formula 
precisely captures the objective of the deobfuscation task. 
The resulting SMT-LIB formula appears in Fig.~\ref{fig:resultingSmtFormula}.

\begin{figure}[htbp]
	\centering
	\begin{adjustbox}{width=0.7\linewidth}
		\begin{minipage}{\linewidth}
			\vspace{0.3em}
			\begin{lstlisting}[label={lst:SMT-LIB-code}]
				(declare-fun ARG_0 () (_ BitVec 32))
				(assert (= (bvsub ARG_0 #x00000003) #00000000))
				(model-add ARG_0 () (_ BitVec 32) #x00000003)
			\end{lstlisting}
		\end{minipage}
              \end{adjustbox}
              \caption{The SMT-LIB formula representing check\_key and
                our RE goal.}
              \label{fig:resultingSmtFormula}
\end{figure}


If we examine the linear equation that corresponds to the assembly code of 
the function \texttt{check\_key(int key)}, and which represents the underlying 
key recovery problem, we obtain the following expression:

\[
EAX - 3 = 0
\]

\mysubsection{Solving the Logical Formula.} The fourth and the last 
step is dispatching the logical formula. In our case-study, this is
performed using the Z3 
SMT-solver, which supports the bit-vector and array theories, and 
 is very efficient in solving these kind of queries. Of course, in
 this case the query is quite easy to solve; but for the highly
 obfuscated programs, the query can grow quite complex. 
In this case, Z3 quickly discovers that the sought-after value is 3.


\section{Deobfuscation}
\label{Deobfuscation_appendix}

\mysubsection{Deobfuscation.}
Existing deobfuscation methods are commonly classified along two 
dimensions:
\emph{static} versus \emph{dynamic} deobfuscation, and \emph{manual} 
versus \emph{automatic} deobfuscation.
\vspace{-1.0em}

\subsubsection{Static VS Dynamic Deobfuscation.}

\emph{Static deobfuscation} involves analyzing and simplifying 
obfuscated code without executing it. This method examines the 
program's binary or source structure to identify applied patterns, 
transformations, and obfuscation techniques. Common tasks include 
locating encrypted or encoded strings, recognizing renamed symbols, 
reconstructing control flow, and rewriting instruction sequences 
into a clearer form. Although static analysis can be effective and 
partially automated, it encounters difficulties with obfuscation 
that depends on runtime behavior, such as dynamic code generation 
or self-modifying code.

\emph{Dynamic deobfuscation} involves executing the code in a controlled 
environment to observe its behavior at runtime. By running the program in 
a sandbox, emulator, or similar setup, the analyst can capture the code 
after runtime transformations --- such as decryption or unpacking --- have taken 
place, revealing details that static analysis cannot observe. This approach 
can expose hidden or conditional code paths, detect advanced techniques 
like code virtualization or dynamic code generation, and provide deeper 
insight into the program's actual functionality.

Key uses of dynamic deobfuscation include:
\vspace{0.1em}
\begin{itemize}
	\item \emph{Runtime decryption} of strings or code segments that become visible only during execution.
	\item \emph{Revealing hidden functionality} that static inspection cannot detect.
	\item \emph{Identifying obfuscation techniques} by monitoring runtime behavior under different conditions.
\end{itemize}
\vspace{-0.9em}

\subsubsection{Manual vs.\ Automated Deobfuscation}

\emph{Manual deobfuscation} is a step-by-step process in which an analyst 
examines the code to identify obfuscation techniques such as meaningless 
names, random or encrypted strings, irregular structure, or suspicious API 
usage. It often includes refactoring code, applying algorithmic 
transformations, or writing custom scripts to undo specific effects.  
Manual analysis may also involve dynamic execution in a controlled 
environment to observe and modify runtime behavior. While powerful, this 
approach is time-consuming and requires significant expertise.

Common indicators of obfuscation include:
\begin{itemize}
	\item Random or encrypted strings.
	\item Nonsensical variable and function names.
	\item Irregular or unnatural code structure.
	\item Unusual program size or disproportionately small code sections.
	\item Use of APIs that dynamically load code or libraries.
\end{itemize}

\emph{Automated deobfuscation} uses specialized tools to simplify or 
recover code with little or no human intervention. Such tools can be 
highly effective against the obfuscation techniques they are designed for, 
but often struggle with unconventional or customized methods. Providing 
fully reliable automated deobfuscation therefore remains challenging.

\section{Case Study: Obfuscated Programs}
\label{Case_Study_appendix}

\begin{figure}[!htbp]
	\centering
	\includegraphics[width=0.9\textwidth, keepaspectratio]{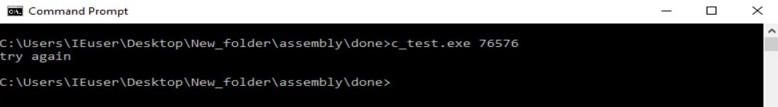}
	\caption{Wrong key}
	\label{fig:failed_execution}
	
	\vspace{1em}
	
	\includegraphics[width=0.9\textwidth, keepaspectratio]{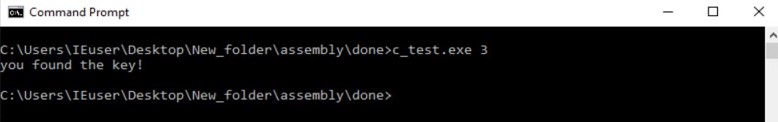}
	\caption{Correct key}
	\label{fig:success_execution}
\end{figure}

\mysubsection{Original Program Execution.} During the execution 
of the original, unobfuscated program, the user is prompted to enter 
a password to ``unlock the locker''. Supplying an incorrect value produces 
a failure message (Fig.~\ref{fig:failed_execution}), whereas entering 
the correct key results in a success message (Fig.~\ref{fig:success_execution}). 
Solving this challenge therefore requires us to reverse engineer the 
program and recover the correct password.

\begin{figure}[!htbp]
	
	\centering
	\begin{subfigure}[t]{0.45\textwidth}
		\includegraphics[width=\linewidth, keepaspectratio]{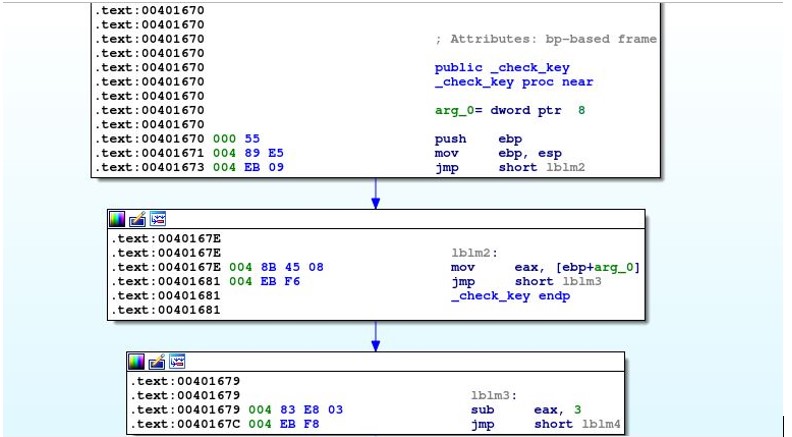}
		\caption{check\_key function --- obfuscated program $P_1$.}
		\label{fig:assembly_p1}
	\end{subfigure}
	
	\centering
	\begin{subfigure}[t]{0.45\textwidth}
		\includegraphics[width=\linewidth, keepaspectratio]{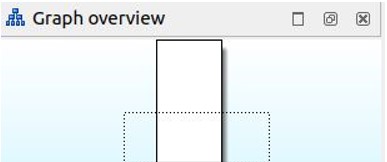}
		\caption{check\_key function --- CFGs of the
                  obfuscated programs $P_2$ and $P_3$.}
		\label{fig:cfg_p2}
	\end{subfigure}
	\hfill
	\begin{subfigure}[t]{0.45\textwidth}
		\includegraphics[width=\linewidth, keepaspectratio]{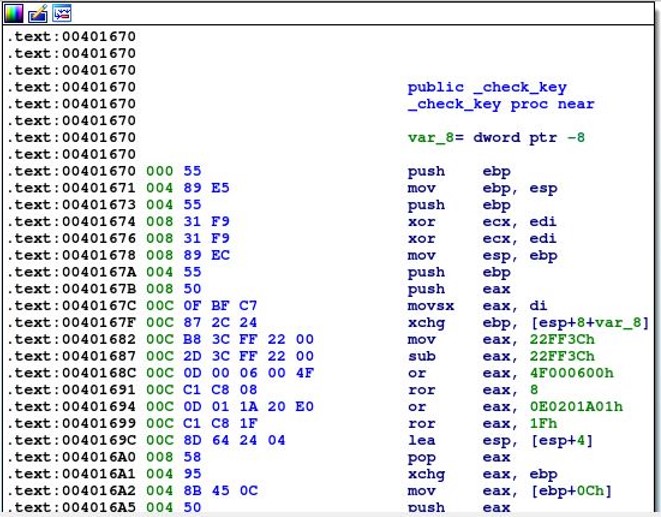}
		\caption{check\_key function --- obfuscated programs
                  $P_2$ and $P_3$.}
		\label{fig:assembly_p2_1}
	\end{subfigure}
	\caption{Obfuscated programs $P_1$, $P_2$ and $P_3$.}
	\label{obfuscated_program_2_3}
	\hfill
	
	\centering
	\begin{subfigure}[t]{0.45\textwidth}
		\includegraphics[width=\linewidth, keepaspectratio]{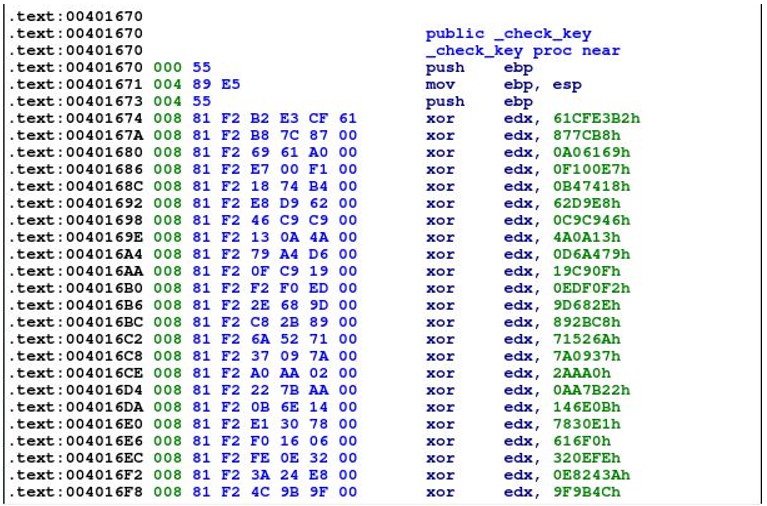}
		\caption{check\_key function --- obfuscated programs $P_4$.}
		\label{fig:assembly_p4_1}
	\end{subfigure}
	\hfill
	\begin{subfigure}[t]{0.45\textwidth}
		\includegraphics[width=\linewidth, keepaspectratio]{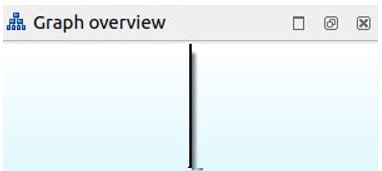}
		\caption{check\_key function --- the CFG of obfuscated
                  program $P_4$.}
		\label{fig:cfg_p4}
	\end{subfigure} 
	\caption{Obfuscated program $P_4$.}
	\label{obfuscated_program_4}  
	\hfill
	
	\centering
	\begin{subfigure}[t]{0.45\textwidth}
		\centering
		\vspace{0pt} 
		\includegraphics[width=\linewidth, keepaspectratio]{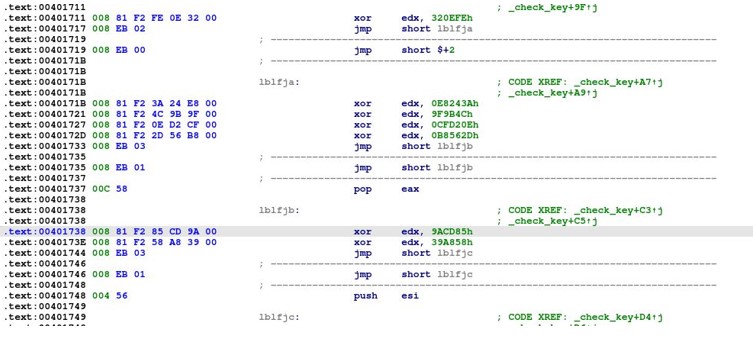}
		\captionsetup{justification=centering}
		\caption{check\_key function --- obfuscated program $P_5$.}
		\label{fig:assembly_p5}
	\end{subfigure}
	\hfill
	\begin{subfigure}[t]{0.45\textwidth}
		\centering
		\vspace{0pt} 
		\includegraphics[width=\linewidth, keepaspectratio]{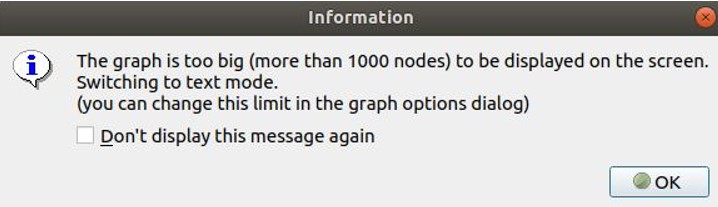}
		\caption{check\_key function --- IDA information
                  message --- cannot display the CFG for $P_5$.}
		\captionsetup{justification=centering}
		\label{fig:warning_msg_ida}
	\end{subfigure} 
	\caption{Obfuscated program $P_5$.}
	\label{obfuscated_program_5}
	
\end{figure}

\mysubsection{The $P_1$--$P_5$ Obfuscated Programs.} In our evaluation 
we examined six programs, $P_0$ through $P_5$, where $P_1$--$P_5$ are 
increasingly obfuscated variants of $P_0$. Figs.~\ref{obfuscated_program_2_3}, 
\ref{obfuscated_program_4}, and \ref{obfuscated_program_5} depict
the obfuscated assembly code and corresponding control-flow graphs 
for these programs.

\section{Running ReSMT}
\label{ReSMT_tool}

The RE property in our case study is to find the correct key; this
means that at the end of the emulation of function
``check\_key(key)'', the ``eax'' register should be equal to zero so
the congratulatory message is printed to the screen.  The RE condition
is stored in a separate Json file that is given as input when running
ReSMT.py. When ReSMT starts, it loads the RE condition
and encodes it into the logical query, after the code emulation phase. The Json
file contains the following relevant fields:
\begin{inparaenum}[(i)]
	\item 
	Input: a list of arguments for the function;
	\item
	Register: the register we want to add a condition on;
	\item
	Operation: the arithmetic operator to be added to the
        condition, from the following options: $==$, $!=$, $<$, $>$; and
	\item
	Value: the integer value to be added.
\end{inparaenum}
An example appears in Fig.~\ref{How_to_run_ReSMT_B}. Finally, the
command line invocation of the tool is depicted in Fig.~\ref{How_to_run_ReSMT_A}.

\begin{figure}[!htbp]
	\centering
	\lstset{basicstyle=\ttfamily\footnotesize, frame=single}    
	\begin{lstlisting}
		{
			"input"		: ["KEY"],
			"register"  : "EAX",
			"operation" : "==",
			"value"     : "0"
		}
	\end{lstlisting}
	\caption{An example Json configuration.}
	\label{How_to_run_ReSMT_B}
	
\end{figure}

\begin{figure}[!htbp]
	\centering
	\lstset{basicstyle=\ttfamily\footnotesize, frame=single}
	\begin{lstlisting}
		python3 ReSMT.py 'binary file' 'function offset' 'function size' 'RE query' 
\end{lstlisting}
        	\caption{Running the Python script.}
	\label{How_to_run_ReSMT_A}
      \end{figure}

\end{document}